# New crystal structure prediction of fully hydrogenated borophene by first principles calculations

Zhiqiang Wang,[1] Tie-Yu Lü,[1] Hui-Qiong Wang,[1,2] Yuan Ping Feng[3], Jin-Cheng Zheng,[1,2,4*]

[1] Department of Physics, and Collaborative Innovation Center for Optoelectronic Semiconductors and Efficient Devices, Xiamen University, Xiamen 361005, China

[2] Xiamen University Malaysia, 439000 Sepang, Selangor, Malaysia

[3] Department of Physics, National University of Singapore, Singapore 117542, Singapore

[4] Fujian Provincial Key Laboratory of Mathematical Modeling and High-Performance Scientific Computation, Xiamen University, Xiamen 361005, China

E-mail: jczheng@xmu.edu.cn

**Abstract**

New crystal structures of fully hydrogenated borophene (borophane) have been predicted by first principles calculation. Comparing with the chair-like borophane (C-boropane) that has been reported in literature, we obtained four new borophane conformers with much lower total-energy. The most stable one, washboard-like borophane (W-borophane), has energy about 113.41 meV/atom lower than C-borophane. In order to explain the relative stability of different borophane conformers, the atom configuration, density of states, charge transfer, charge density distribution and defect formation energy of B-H dimer have been calculated. The results show that the charge transfer from B atoms to H atoms is crucial for the stability of borophane. In different borophane conformers, the bonding characteristics between B and H atoms are similar, but the B-B bonds in W-borophane are much stronger than that in C-borophane or other structures. In addition, we examined the dynamical stability of borophane conformers by phonon dispersions and found that the four new conformers are all dynamically stable. Finally the mechanical properties of borophane conformers along an arbitrary direction have been discussed. W-borophane possesses unique electronic structure (Dirac cone), good stability and superior mechanical properties. W-borophane has broad perspective for nano electronic device.

**Keywords**：Borophane; Stability; Conformers; First principles calculations

**Introduction**

Two-dimensional (2D) boron sheet (borophene) was synthesized on a silver substrate under ultrahigh-vacuum[1] and has attracted much attention[2-19]. Difference from graphene which is an isotropic material, borophene shows high anisotropy due to the different arrangements of B atoms along the *a* and *b* direction. It has been shown that the mechanical properties of borophene are highly anisotropic[8]. Furthermore, the electronic and transport properties are also strongly dependent on the direction.[20] The superconducting transition temperature $T_c$ of borophene is about 19 K, which can be increased to 27.4 K by applying a tensile strain, or 34.8 K by hole doping[9], similar with previous study on strained $MgB_2$[21]. The lattice thermal conductivity of borophene is unexpectedly low on account of the strong phonon-phonon scattering[22]. In addition, borophene shows vast potentials as an anode material for Li, Na and Mg ion batteries due to high theoretical specific capacities and outstanding ion transport properties.[18, 23-26]

Borophene is unstable against long-wavelength transversal waves[27] due to the small imaginary frequency along the Γ-X direction in the phonon dispersion. It has been reported that full hydrogenation can stabilize borophene. After full hydrogenation, no imaginary frequencies were found along the high-symmetry directions of the Brillouin zone. The fully hydrogenated borophene has a direction-dependent Dirac cone, and the Dirac fermions possess an ultrahigh Fermi velocity ($3.0 \times 10^6$ m/s)[27]. This, combined with the excellent mechanical performance, makes fully hydrogenated borophene attractive for applications in nanoelectronics devices. Hydrogenation is an important approach to modify the physical and chemical properties of 2D materials. Different hydrogenation patterns and coverage on the same 2D material can lead to different physical and chemical properties[28-30], including the band structure[31-34], optical properties[35], magnetic properties[36] and mechanical properties[37]. Furthermore, under the same hydrogenation coverage, the mechanical properties are also strongly dependent on the atomic configuration.[38] It is important to explore stable configurations of borophane. To date, conformers of borophane other than C-borophane have not been reported, and would be the focus of the present study.

In this work, the structural stability, band structures, charge density distribution and mechanical properties of borophanes with different configurations have been studied by first principles calculations. We obtained four new conformers with much lower total-energy than that of C-borophane. Furthermore, we established the relative stability of different borophane conformers. By analyzing the atom configurations and the total-energies of borophane conformers, we found the configuration in which B atoms are staggered by zigzag mode along the *a* direction, and staggered by up and down wrinkle mode along the *b* direction is the most stable one. Furthermore, we explain the stability of borophane conformers by analyzing density of states, charge transfer, charge density distribution and defect formation energy of B-H dimer. The results show that the charge transfer from B atoms to H atoms is crucial for the stability of borophane. The charge density distribution and defect formation energy of B-H dimer show that the B-B bonds in W-borophane are stronger than that in C-borophane. Moreover, we examined the dynamical stability of borophane conformers by calculating the phonon dispersions. Finally the electronic structures and the mechanical properties of the C-borophane and the four new conformers have been discussed.

**Computational details**

All calculations are performed using the Quantum-Espresso package[39]. Ultrasoft pseudopotentials[40] are used for all atoms and the exchange-correlation approximation is evaluated through the Perdew-Burke-Ernzerh (PBE)[41] functional. The kinetic-energy cutoff of plane-waves is set to be 50 Ry. The atomic positions and lattice constants are fully relaxed, until the forces on all atoms are less than 0.01 eV/Å. A large vacuum region (20 Å) is included along $z$ direction to eliminate the interlayer interactions.

The four nonzero elastic constants of a 2D material are $c_{11}$, $c_{22}$, $c_{12}$ and $c_{66}$. The Young's modulus along the $a$ and $b$ directions can be written as[42]

$$Y_a = \frac{c_{11}c_{22} - c_{12}^2}{c_{22}} \quad \text{and} \quad Y_b = \frac{c_{11}c_{22} - c_{12}^2}{c_{11}}, \qquad 1$$

respectively, the Poisson's ratio along the $a$ and $b$ directions can be given as

$$v_a = \frac{c_{12}}{c_{22}} \quad \text{and} \quad v_b = \frac{c_{12}}{c_{11}}, \qquad 2$$

respectively, while the shear modulus can be expressed as

$$G = c_{66}. \qquad 3$$

**Results and discussion**

**Crystal structures and stability**

In order to study the influence of H adsorption on the atomic and electronic structure of borophene, we calculated the borophene with a single H atom adsorption in the 6×2×1 supercell. Four different adsorption sites have been taken into consideration. Our results show that the most stable adsorption site is the top site of B atom. Significant shifts of B atoms have been observed when the H atom is adsorbed on the top site. The local atomic structure schematic of borophene with one, two and three H atoms adsorption have been shown in Figure 1. The shifts of B atoms and the average adsorption energy of H atoms are listed in Table 1. As shown in Figure 1 (a), in the single H atom adsorption, the shift of the B atom (B3) bonded with the H atom is about 0.32 Å along Z direction. For the two first neighbor B atoms (B2 and B4) of the B atom bonded with the H atom, the shifts along the negative Z direction are both about 0.3 Å. However, for the second and third neighbor B atoms of B3, the shifts along z direction are only 0.04 and 0.03 Å, indicating that H atom induced atomic distortion is mainly imposed to the first neighbor B atom of the B atom bonded with H atom. For two H atoms adsorbed borophene, two configurations have been shown. As shown in Figure 1 (b), two H atoms adsorbed on two neighbor B atoms. Adsorption of H atom on B2 atom will lead the shift of B3 atom along negative Z direction. However, adsorption of H atom on B3 atom will make B3 atom move along positive Z direction. For B3 atom, the atomic shifts induced by the two adsorbed H atoms are not along the same direction. However, in H2-B (Figure 1 (c)), for B3 atom, the atomic distortion induced by the two adsorbed H atoms are consistently, leading to that the average adsorption energy of H atom in H2-B configuration is 0.38 eV/H higher than that in H2-A. For three H atoms adsorption, in H3-A, three H atoms are adsorbed on the three neighbor B atoms, while, in H3-B, three H atoms are adsorbed on the three next neighbor B atoms. Similarly, the configuration H3-B is much more stable than H3-A. More specifically, the average adsorption energy of H atoms in H3-B configuration is 0.65 eV/H higher than that in H3-A. In addition, by analyzing the partial density of states of H and B atoms in H3-A and H3-B, we found that the orbital hybridization between H (*s*) and B (*p$_z$*) in H3-B is much stronger than that in H3-A.

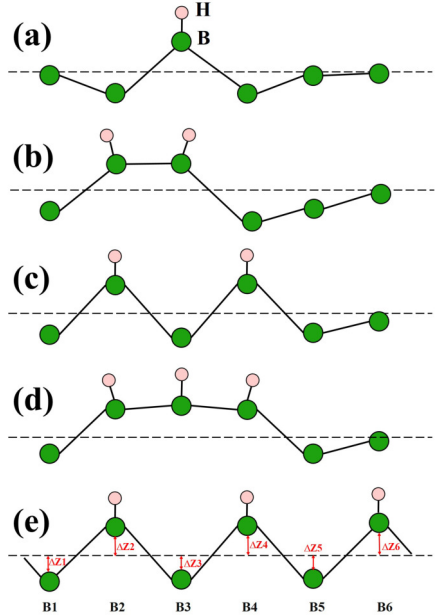

Figure 1. The local atomic structure schematic of borophene with one, two and three H atoms adsorption. (a) H1, (b) H2-A, (c) H2-B, (d) H3-A and (e) H3-B.

Table 1. The shifts of B atoms along Z direction and average adsorption energy of H atoms in H1, H2-A, H2-B, H3-A and H3-B. When ΔZ>0, the corresponding B atom shifts along positive Z direction. When ΔZ<0, the corresponding B atom shifts along negative Z direction. The adsorption energy $E_{ad}$ can be written as $E_{ad} = (E_{\text{borophene}} + nE_{\text{H}} - E_{\text{borophene-H}})/n$, where $E_{\text{borophene}}$, $E_{\text{H}}$, $E_{\text{borophene-H}}$ and $n$ are the total energy of borophene, isolated H atom, borophene with H atoms adsorption and the number of H atoms.

| Configurations | ΔZ1 (Å) | ΔZ2 (Å) | ΔZ3 (Å) | ΔZ4 (Å) | ΔZ5 (Å) | ΔZ6 (Å) | $E_{ad}$ (eV) |
|---|---|---|---|---|---|---|---|
| H1 | -0.04 | -0.30 | 0.32 | -0.30 | -0.04 | 0.03 | **3.95** |
| H2-A | -0.38 | 0.32 | 0.31 | -0.39 | -0.09 | -0.09 | 3.60 |
| H2-B | -0.27 | 0.26 | -0.48 | 0.25 | -0.27 | 0.03 | **3.98** |
| H3-A | -0.50 | 0.24 | 0.32 | 0.24 | -0.74 | -0.48 | 3.53 |
| H3-B | -0.50 | 0.20 | -0.50 | 0.20 | -0.50 | 0.20 | **4.18** |

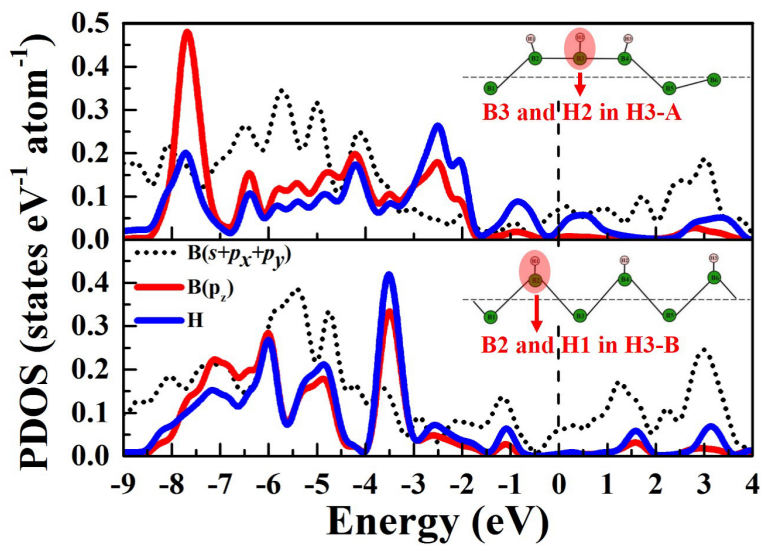

Figure 2. Partial density of states of B atoms and H atoms in three H atoms adsorbed borophene.

Then we studied the atomic, electronic structures and stability of fully hydrogenated borophene. The crystal structures of seven borophane conformers are displayed in Figure 3. For convenience of description, we named the seven conformers as Plane-Square-type borophane (PS-borophane), Plane-Triangle-type borophane (PT-borophane), Chair-like borophane, Boat-like borophane (B-borophane), Twist-Chair-Boat-type borophane (TCB-borophane), Triangle-type borophane (T-borophane) and Washboard-like borophane, respectively. The unit cells are marked by the black dashed rectangles. The optimized lattice constants, B-H bond lengths, buckling heights and the total-energy difference relative to C-borophane are listed in Table 2. In the seven borophane conformers, each B atom is hydrogenated with an H atom, namely, the ratio of B:H is 1:1. The C-borophane configuration has been reported, previously [20, 27, 43]. As shown in Figure 3(e) and (f), C-Borophane has a buckled configuration and the buckling height is 0.80 Å, which is smaller than that of borophene (0.91 Å)[8]. In C-borophane, Hydrogen atoms alternate on both sides of the borophene sheet. B atoms are aligned along the *a* direction and staggered by up and down wrinkle mode along the *b* direction. For comparison, two planner conformers (PS- and PT-borophane) that all hydrogen atoms bond with B atoms at the same side have been taken into consideration. In PS-borophane, each B atom bonds with four neighboring B atoms. In other six conformers, each B atom bonds with six neighboring B atoms. In PS-borophane, all B atoms are aligned along the *a* and *b* direction. In PT-borophane, B atoms are aligned along the *a* direction, however, staggered by zigzag mode along the *b* direction. The total-energy difference relative to that of C-borophane and the schematic drawings of different conformers are shown in Figure 4. PS- and PT-borophane have total-energy difference 685.93 and 425.67 meV/atom higher than C-borophane, respectively, indicating that the buckled configuration is more stable than the planner configurations.

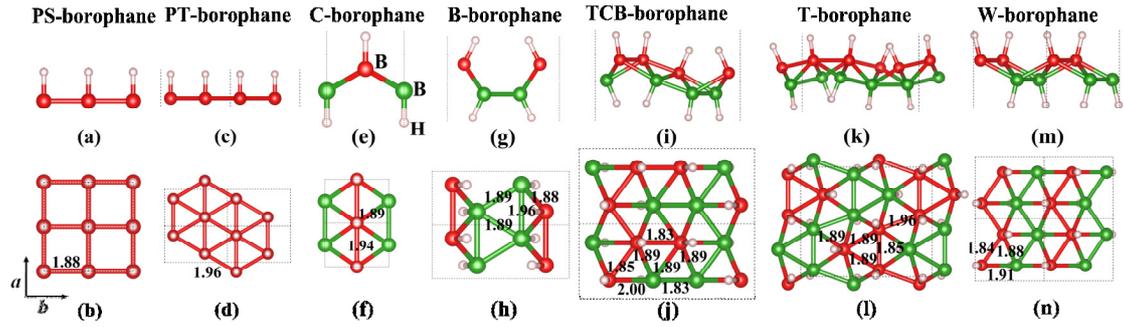

Figure 3. Crystal structure of seven borophane conformers. (a) side view with *b-c* plane shown, (b) top view of atomic structure of PS-borophane. (c) side view with *b-c* plane shown, (d) top view of atomic structure of PT-borophane. (e) side view with *b-c* plane shown, (f) top view of atomic structure of C-borophane. (g) side view with *b-c* plane shown, (h) top view of atomic structure of B-borophane. (i) side view with *b-c* plane shown, (j) top view of atomic structure of TCB-borophane. (k) side view with *b-c* plane shown, (l) top view of atomic structure of T-borophane. (m) side view with *b-c* plane shown, (n) top view of atomic structure of W-borophane. The numbers are the B-B bond lengths in Å. The big red and green balls represent B atoms, the small light pink balls represent H atoms, respectively.

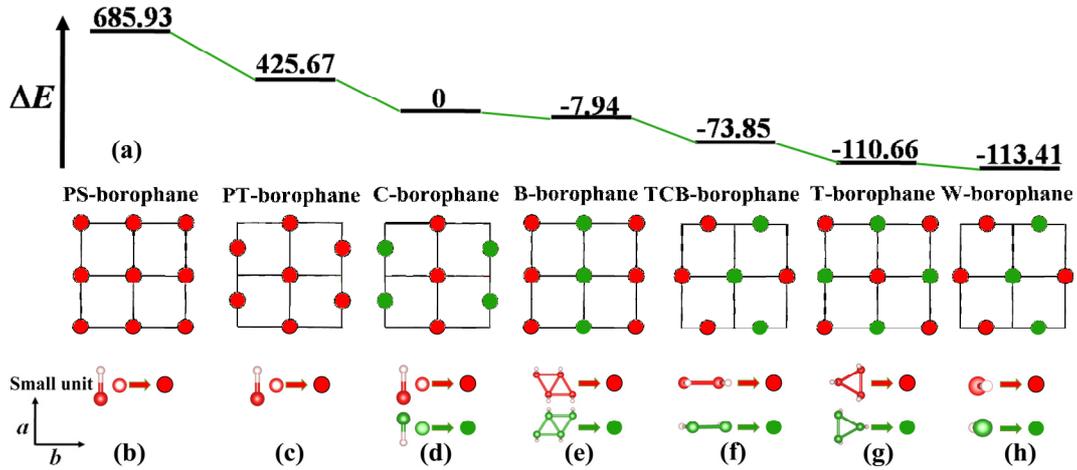

Figure 4. (a) The total-energy difference that referred to the total-energy C-borophane of different borophane conformers. Schematic drawings of the small units arrangements in (b) PS-, (c) PT-, (d) C-, (e) B-, (f) TCB-, (g) T- and (h) W-borophane.

Table 2. The lattice constants, B-H bond lengths $d_{B-H}$, buckling heights $d_{Buckling}$ and the total-energy difference relative to C-borophane at PBE, PW91 and LDA level of PS-, PT-, C-, B-, TCB-, T- and W-borophane.

| Configurations | $a$ (Å) | $b$ (Å) | $d_{B-H}$ (Å) | $d_{Buckling}$ (Å) | $\Delta E$ (meV/atom) | | |
|---|---|---|---|---|---|---|---|
| | | | | | PBE | PW91 | LDA |
| PS-borophane | 1.876 | 1.876 | 1.210 | 0.00 | 685.93 | 684.72 | 672.22 |
| PT-borophane | 1.966 | 3.392 | 1.179 | 0.00 | 425.67 | 427.56 | 417.07 |
| C-borophane | 1.941 | 2.815 | 1.178 | 0.81 | 0.00 | 0.00 | 0.00 |
| B-borophane | 1.955 | 5.013 | 1.180 | 1.34 | -7.94 | -8.00 | -13.66 |
| TCB-borophane | 2.969 | 6.933 | 1.186 | 1.61 | -73.85 | -73.62 | -82.87 |

| T-borophane | 5.010 | 6.009 | 1.190 | 1.24 | -110.66 | -110.18 | -124.63 |
| W-borophane | 3.039 | 3.379 | 1.190 | 0.89 | -113.41 | -112.97 | -127.79 |

C-, B-, TCB-, T- and W-borophane are all buckled configurations. In B-borophane, for convenience of description, we regard the four neighboring B atoms that constitute a parallelogram and the four H atoms that bond with the four B atoms at the same side as a small unit, as shown in Figure 4 (e). The small units are aligned along the *a* direction, however, staggered by up and down wrinkle mode along the *b* direction. B-borophane has energy 7.94 meV/atom lower than C-borophane. In TCB-borophane, pairs of hydrogen atoms alternate on both sides along the *a* and *b* direction. Similarly, we regard the two neighboring B atoms that bond with two H atoms at the same side and the two H atoms as a small unit, as shown in Figure 4 (f). The small units are staggered by zigzag mode along the *a* direction, and staggered by up and down wrinkle mode along the *b* direction. TCB-borophane has energy 73.85 meV/atom lower than C-borophane. In T-borophane, ternate hydrogen atoms alternate on both sides along the *a* and *b* direction. We regard the three neighboring B atoms that constitute a triangle and the three H atoms that bond with the three B atoms at the same side as a small unit, as shown in Figure 4 (g). The small units are staggered by up and down wrinkle mode along the *a* and *b* direction. T-borophane has energy 110.66 meV/atom lower than C-borophane. In W-borophane, B atoms are staggered by zigzag mode along the *a* direction, and staggered by up and down wrinkle mode along the *b* direction. W-borophane has energy difference about 113.41 meV/atom lower than C-borophane. Compared with C-borophane (wrinkle along *b* direction), W-borophane has buckled configuration both along *a* and *b* direction, leading to superior mechanical properties along *a* direction. Specifically, the ultimate tensile strain along *a* direction of W-borophane is 0.17, which is larger than that of C-borophane (0.12). In order to further affirm the relative stability of different borophane conformers, different exchange-correlation functionals with local density approximation (LDA) and general gradient approximation (GGA, such as typical PBE, and PW91) have been used to check the trend of the total-energy differences. PBE and PW91 are both belong to GGA exchange-correlation functional, therefore, the energy difference (ΔE) obtained by PBE and PW91 are highly consistent. The overestimation of the cohesive energy at the LDA level has been reported[44], resulting in the distinct difference of ΔE at the LDA and GGA level. On the whole, the trends of energy differences obtained by the three exchange-correlation functionals are consistent. This further indicates that the stability of the new borophene conformers is independent or less dependent on the form of exchange-correlation functionals. It is interesting that the energy difference between T- and W-borophane is only 2.75 meV/atom under the stress-free states. Under ≥0.05 biaxial tensile stains, the total-energy of T-borophane change to be lower than that of W-borophane, indicating that there exists a structural phase transition between T- and W-borophane under increasing biaxial tensile strains.

In order to examining the dynamical stability of borophane conformers, we calculated the phonon dispersions. Phonon dispersion is an important parameter to estimate the dynamical stability of crystal structure. Imaginary frequencies along any high-symmetry direction of the Brillouin zone are indications of dynamic instability of the crystal structure.

The calculated phonon dispersions of B-, TCB-, T- and W-borophane are shown in Figure 5. For the four new borophane conformers, no imaginary frequencies were found along the high-symmetry directions of the Brillouin zone, indicating that the four new borophane conformers are all dynamically stable.

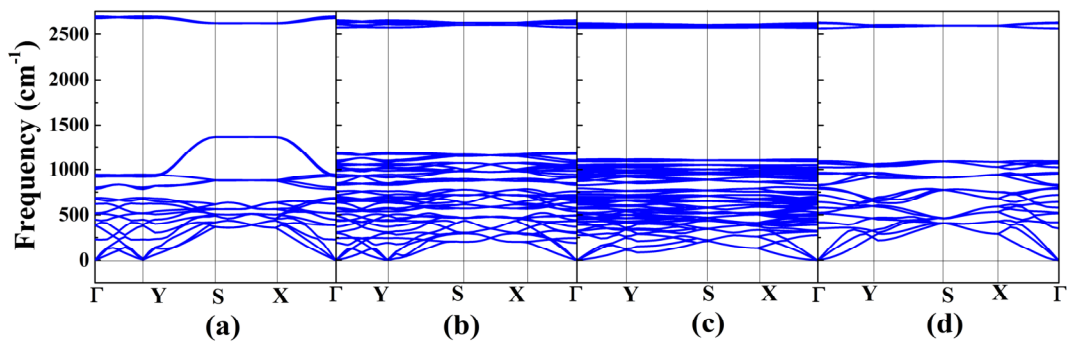

Figure 5. The phonon dispersions of (a) B-, (b) TCB-, (c) T- and (d) W-borophane.

**Electronic structures**

In order to further explain the excellent stabilities of W-borophane, we analyzed the electronic structures of different borophane conformers to obtain a deep insight of the bonding characteristics and stabilization mechanism. A simple and clear picture of electronic bonding in boron sheet has been reported[45]. A structure that optimally fills in-plane bonding states ($s$, $p_x$ and $p_y$) should be most preferable[45]. In borophene, in-plane $sp^2$ antibonding states are partly occupied. Hence, borophene is unstable and prone to donate electrons[46]. In hydrogenated borophene, there exist a charge transfer from B atoms to H atoms, leading to that the in-plane bonding states are completely filled and the antibonding states are empty; moreover, the out-of-plane bonding states are also fully filled. Finally, the Fermi level is exactly located at the valley bottom of the total density of states. Consequently, full hydrogenated borophene is stable. The band structures and density of states of B-, TCB-, T- and W-borophane are displayed in Figure 6. The four new borophane conformers all possess a Dirac cone. The band structure of C-borophane shows clearly a Dirac cone along the Γ-X direction[27,43]. Similarly, the band structure of B-borophane shows a Dirac cone along the Γ-X direction. However, for TCB-, T- and W-borophane, the Dirac cones are along the Γ-Y direction. Furthermore, in order to analyze the electronic bonding of W-borophane, we calculated the partial density of states (PDOS) of H and B atoms, as shown in Figure 7. By analyzing the energy rang and peak position, we found that there exists a strong orbital hybridization between $s$ orbit of H atoms and $p_z$ orbit of B atoms. Furthermore, from the partial density of states near the Fermi level, we can found that the Dirac electron is mainly contributed by the $p_x$ and $p_y$ electron of B atoms for W-borophane. A similar phenomenon has been observed in C-borophane.[27].

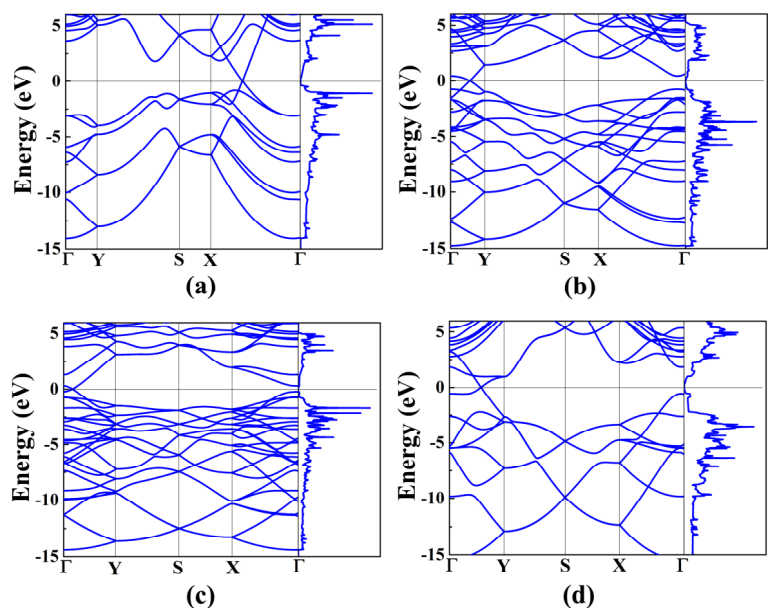

Figure 6. Band structures and density of states of (a) B-, (b) TCB-, (c) T- and (d) W-borophane.

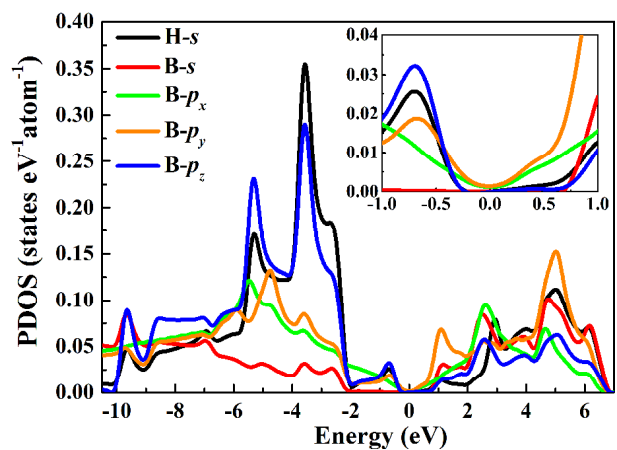

Figure 7. Partial density of states of H and B atoms in W-borophane.

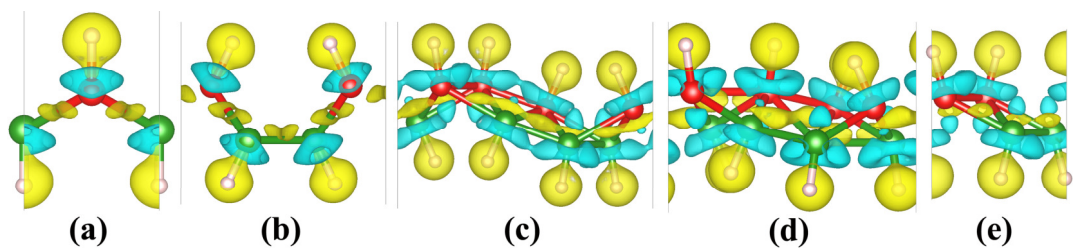

Figure 8. Deformation charge density of (a) C-, (b) B-, (c) TCB-, (d) T- and (e) W-borophane.

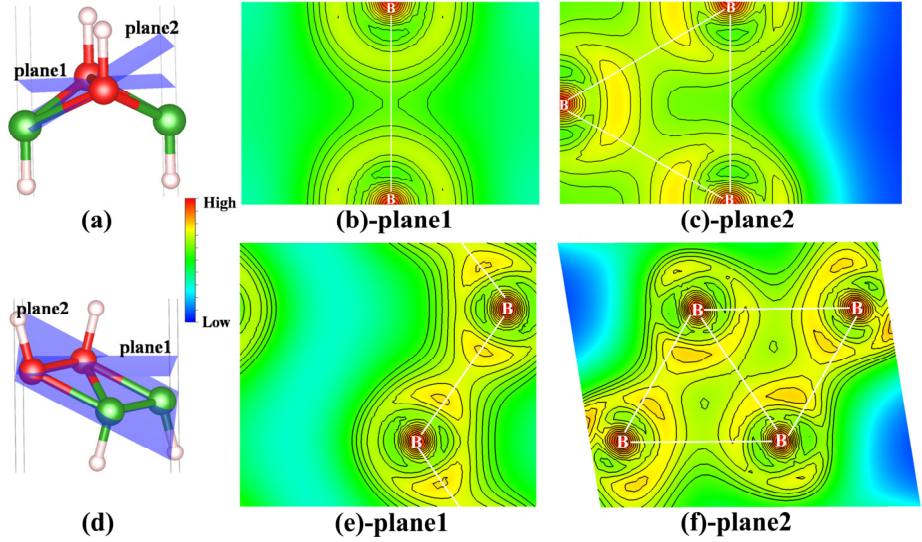

Figure 9. 2D charge density contour of two typical planes in C- and W-borophane. Plane1 is perpendicular to *z* coordinate axis and contains a B-B bond. Three nearest neighboring B atoms that constitute a triangle determine plane2. The red area represents the electron is dense in the area, however, the blue area represents the electron is sparse in the area.

In order to confirm the aforementioned charge transfer, we calculated the deformation charge density of C-, B-, TCB-, T- and W-borophane. The deformation charge density is defined as the difference between the total charge density ($\rho(\vec{r})$) in the solid and the superposition of independent atomic charge densities ($\sum_i \rho_{atom}(\vec{r} - \vec{r}_i)$) placed at the atomic sites of the same solid, the equation can be written as

$$\Delta \rho(\vec{r}) = \rho(\vec{r}) - \sum_i \rho_{atom}(\vec{r} - \vec{r}_i) \quad . \qquad 4$$

The yellow areas represent electron accumulation ($\Delta\rho(\vec{r}) > 0$), and the blue areas represent electron depletion ($\Delta\rho(\vec{r}) < 0$). As shown in Figure 8, a charge transfer from B atoms to H atoms is clearly shown. By comparing the electronegativity of boron and hydrogen, we can find that the electronegativity of hydrogen (2.2) is larger than that of boron (2.04). Therefore, the charge transfer from B atoms to H atoms is reasonable. Similarly, a charge transfer from silicon (1.9) to hydrogen (2.2) atom has been found in hydrogenated silicene.[47] However, the electronegativity of carbon (2.55) is larger than that of hydrogen (2.2), on the contrary, a charge transfer from H atoms to C atoms has been observed in graphane.[32]

Furthermore, we investigated the bonding characteristics in C- and W-borophane by analyzing the charge density distributions and defect formation energy of a B-H dimer. The in-plane bonds (B-B bonds) are stronger that out-of-plane π-bonds (B-H bonds). In PT-, C-, B-, TCB-, T- and W-borophane, the B-H bond lengths are almost the same, indicating that the bonding characteristics between B and H atoms are uniform. Similar phenomenon has been reported in different conformers of hydrogenated silicene[47]. Hence, it is imperative to analyze the bonding characteristics between B and B atoms in borophane conformers. The 2D charge density contours of C- and W-borophane are shown in Figure 9. We have selected two typical planes in C- and W-borophane. Plane1 is perpendicular to *z* coordinate axis and contains a B-B bond. Three nearest neighboring B atoms that constitute a triangle determine plane2. Comparing the charge density distribution of C- with W-borophane, we can find that more electrons gathered along the B-B bond direction in W-borophane than that in C-borophane, indicating that the B-B bonds in W-borophane are stronger than that in C-borophane. In order to quantitative describe the bonding strength of B-B bonds, we calculated the defect formation energy of a B-H dimer in C- and W-borophane. The defect formation energy $E_{form}$ can be expressed as

$$E_{form} = E_{def} - \frac{N_{def}}{N_{per}} E_{per},\qquad 5$$

where $E_{def}/E_{per}$ is the total energy of the supercell with/without the B-H dimer defect, and $N_{def}/N_{per}$ denote the atom numbers of the supercell with/without the B-H dimer defect. As shown in Figure 10, when the defect concentration η=1/24, the defect formation energy of a B-H dimer is only 0.40 eV for C-borophane, however, that value change to 0.95 eV for W-borophane. The defect formation energy of B-H dimer in W-borophane is much higher than that in C-borophane, indicating that the B-B bonds in W-borophane are much stronger than that in C-borophane. It is agree well with the results of charge density distributions.

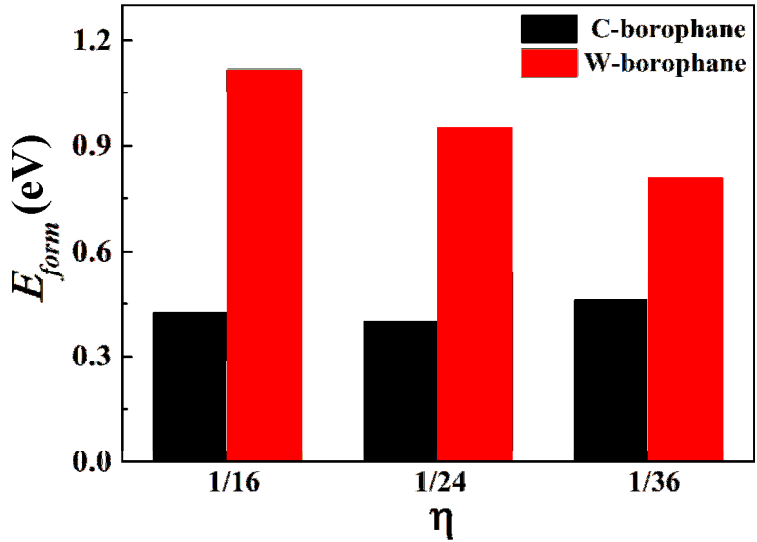

Figure 10. Defect formation energy of a B-H dimer in C- and W-borophane. One B-H dimer defect has been introduced into the supercell, hence, defect concentration η can be expressed as η=1/$N_{per}^{B}$, where $N_{per}^{B}$ is the number of B atoms in the perfect supercell.

**Mechanical properties**

In addition, we calculated the elastic constants, Young's modulus, shear modulus, and Poisson's ratios of B-, TCB-, T- and W-borophane. As listed in Table 3, the Young's modulus of C-borophane along the *a* and *b* direction are 172.24 and 110.59 N/m[43]. For C- and B-borophane, the Young's modulus along the *a* direction is much larger than that along the *b* direction. By analyzing the atomic structure of C- and B-borophane, we found that B atoms are aligned along the *a* direction, however, staggered by up and down wrinkle mode along the *b* direction. Under uniaxial tensile stains along the *a* direction, the B-B bonds along the *a* direction were elongated, significantly. However, the uniaxial tensile strains along the *b* direction can be released due to the buckled configuration. For C- and B-borophane, the ratio ($Y_a/Y_b$) of Young's modulus along the *a* and *b* direction are 1.56 and 2.37, respectively. However, the ratio is reduced to 1.13 for W-borophane. Similarly, we found that all B atoms in W-borophane are staggered along the *a* and *b* direction, therefore, the uniaxial strains along the *a* and *b* direction can be released. In addition, we also calculated the mechanical properties of B-, TCB-, T- and W-borophane along an arbitrary direction. The results are shown in Figure 11. For isotropic materials, the Young's modulus and shear modulus are independent of the direction. The polar diagrams of Young's modulus and shear modulus are perfect circles. The larger degree of deviation from a perfect circle, the stronger anisotropy the materials possess. The Young's modulus and shear modulus of C-borophane are strongly dependent on the direction[43]. However, for W-borophane, the Young's modulus and shear modulus tend to more isotropic.

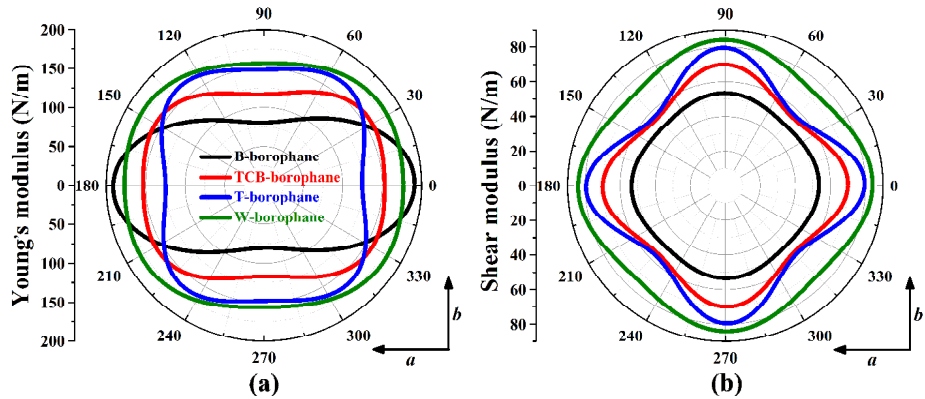

Figure 11. Polar diagrams for Young's modulus and shear modulus of B-, TCB-, T- and W-borophane. The angle is measured relative to the *a* direction. Isotropic (anisotropic) behavior is associated to a circular (noncircular) shape.

**Table 3**. Elastic constants $c_{ij}$, shear modulus $G$, Young's modulus $Y$ in N/m, and Poisson's ratio $v$ of C-, B-, TCB-, T- and W-borophane.

| Configurations | $c_{11}$ | $c_{22}$ | $c_{12}$ | $c_{66}$=G | $Y_a$ | $Y_b$ | $v_a$ | $v_b$ |
|---|---|---|---|---|---|---|---|---|
| C-borophane[43] | 175.77 | 112.86 | 19.97 | 28.46 | 172.24 | 110.59 | 0.177 | 0.144 |
| B-borophane | 197.00 | 83.00 | 22.00 | 53.50 | 191.17 | 80.54 | 0.265 | 0.112 |
| TCB-borophane | 160.00 | 122.00 | 29.00 | 70.00 | 153.11 | 116.74 | 0.238 | 0.181 |
| T-borophane | 129.50 | 154.00 | 27.50 | 79.50 | 124.59 | 148.16 | 0.179 | 0.212 |
| W-borophane | 179.50 | 159.00 | 21.00 | 84.00 | 176.73 | 156.54 | 0.132 | 0.117 |

**Conclusions**

In summary, with first principles calculations we have studied the structure stability, electronic structures and mechanical properties of borophane with different configurations. Comparing with C-boropane, we obtained four new conformers with much lower total-energy. By analyzing the atomic arrangements and the total energies in different conformers, we found that the configuration that B atoms are staggered by zigzag mode along the *a* direction and staggered by up and down wrinkle mode along the *b* direction is the most stale one. The most stable one, W-borophane, has energy difference about 113.41 meV/atom lower than C-borophane. The charge transfer from B atoms to H atoms is crucial for the stability of borophane. Furthermore, the results of charge density distribution show that more electrons gathered along the B-B bond direction in W-borophane than that in C-borophane, moreover, the defect formation energy of B-H dimer in W-borophane is much higher than that in C-borophane, both indicating that the B-B bonds in W-borophane are stronger than that in C-borophane. By calculating the phonon dispersions of the four new borophane conformers, we found no imaginary frequencies along the high-symmetry directions of the Brillouin zone, indicating that the four new conformers are all dynamically stable. The band structures of the four new conformers all show a Dirac cone along Γ-Y or Γ-X direction. Furthermore, from the partial density of states near the Fermi level, we can found that the Dirac electron is mainly contributed by the $p_x$ and $p_y$ electron of B atoms. The unique electronic structure results in high carrier mobility, making borophane a promising material for nano electronic device. Finally the mechanical properties of borophane conformers along an arbitrary direction have been discussed. The results show that the Young's modulus and shear modulus of C-borophane are highly anisotropic. However, for W-borophane, the Young's modulus and shear modulus tend to more isotropic.

**Acknowledgements**
This work is supported by the Fundamental Research Funds for Central Universities (Grant Nos. 20720160020, 2013121010), the Natural Science Foundation of Fujian Province, China (Grant No. 2015J01029), Special Program for Applied Research on Super Computation of the NSFC-Guangdong Joint Fund (the second phase), the National Natural Science Foundation of China (nos. 11335006, 51661135011). TYL acknowledges the National University of Singapore for hosting his visit during which part of the work reported here was carried out.

**Author Contributions**
ZW performed the calculation and drafted the manuscript. HQW and TYL did the data analysis and revised the manuscript. YPF participated in the discussions and revised the manuscript. JCZ proposed and led the project and revised the manuscript.

 **Competing financial interests:** The authors declare no competing financial interests.